\documentclass[12pt,oneside,reqno]{amsart}
\usepackage{amsmath,amssymb,amsthm,textcomp}
\usepackage{amsfonts,graphicx}
\usepackage[mathscr]{eucal}
\usepackage{color}
\usepackage[table]{xcolor}
\usepackage{booktabs}
\usepackage{epstopdf}

\usepackage{subfig}
\usepackage{epsfig}
\usepackage{caption}
\usepackage{hyperref}
\theoremstyle{definition}
\usepackage{float}
\usepackage{multirow}
\floatstyle{plaintop}
\restylefloat{table}
\usepackage{lscape}
\usepackage[T1]{fontenc}
\usepackage[utf8]{inputenc}
\usepackage{mathtools}

\newcommand{\ncom}{\newcommand}

\ncom{\beq}{\begin{equation}}
\ncom{\eeq}{\end{equation}}
\ncom{\bea}{\begin{eqnarray*}}
	\ncom{\eea}{\end{eqnarray*}}
\ncom{\beqa}{\begin{eqnarray}}
\ncom{\eeqa}{\end{eqnarray}}
\ncom{\nno}{\nonumber}
\ncom{\non}{\nonumber}
\ncom{\ds}{\displaystyle}
\ncom{\half}{\frac{1}{2}}
\ncom{\mbx}{\makebox{.25cm}}
\ncom{\hs}{\mbox{\hspace{.25cm}}}
\ncom{\rar}{\rightarrow}
\ncom{\Rar}{\Rightarrow}
\ncom{\noin}{\noindent}
\ncom{\bc}{\begin{center}}
	\ncom{\ec}{\end{center}}
\ncom{\sz}{\scriptsize}
\ncom{\rf}{\ref}
\ncom{\s}{\sqrt{2}}
\ncom{\sgm}{\sigma}
\ncom{\Sgm}{\Sigma}
\ncom{\psgm}{\sigma^{\prime}}
\ncom{\dt}{\delta}
\ncom{\Dt}{\Delta}
\ncom{\lmd}{\lambda}
\ncom{\Lmd}{\Lambda}
\ncom{\Th}{\Theta}
\ncom{\e}{\eta}
\ncom{\eps}{\epsilon}
\ncom{\pcc}{\stackrel{P}{>}}
\ncom{\lp}{\stackrel{L_{p}}{>}}
\ncom{\dist}{{\rm\,dist}}
\ncom{\sspan}{{\rm\,span}}
\ncom{\re}{{\rm Re\,}}
\ncom{\im}{{\rm Im\,}}
\ncom{\sgn}{{\rm sgn\,}}
\ncom{\ba}{\begin{array}}
	\ncom{\ea}{\end{array}}
\ncom{\hone}{\mbox{\hspace{1em}}}
\ncom{\htwo}{\mbox{\hspace{2em}}}
\ncom{\hthree}{\mbox{\hspace{3em}}}
\ncom{\hfour}{\mbox{\hspace{4em}}}
\ncom{\vone}{\vskip 2ex}
\ncom{\vtwo}{\vskip 4ex}
\ncom{\vonee}{\vskip 1.5ex}
\ncom{\vthree}{\vskip 6ex}
\ncom{\vfour}{\vspace*{8ex}}
\ncom{\norm}{\|\;\;\|}
\ncom{\integ}[4]{\int_{#1}^{#2}\,{#3}\,d{#4}}
\ncom{\vspan}[1]{{{\rm\,span}\{ #1 \}}}
\ncom{\dm}[1]{ {\displaystyle{#1} } }
\ncom{\ri}[1]{{#1} \index{#1}}
\newtheoremstyle
{remarkstyle}
{}
{11pt}
{}
{}
{\bfseries}
{:}
{     }
{\thmname{#1} \thmnumber{#2} }

\theoremstyle{remarkstyle}

\usepackage{setspace}

\def\eps{\varepsilon}

\def\E{{\mathbb E}}

\begin{document}
	\title{\large D\lowercase{iscussion}  \lowercase{on} `A \lowercase{new}  \lowercase{extension} \lowercase{of} \lowercase{the} FGM \lowercase{copula} \lowercase{with} \lowercase{application} \lowercase{in} \lowercase{reliability}'}
	\author[Ashok Kumar Pathak]{Ashok Kumar Pathak$^{1}$}
	\author{Mohd. Arshad$^{2, *}$
			\\\\ 
		$^{1}$D\lowercase{epartment of} M\lowercase{athematics and} S\lowercase{tatistics}, C\lowercase{entral} U\lowercase{niversity of} P\lowercase{unjab},
		B\lowercase{athinda}, I\lowercase{ndia}.	
		\\$^{2}$D\lowercase{epartment of} M\lowercase{athematics}, I\lowercase{ndian} I\lowercase{nstitute of} T\lowercase{echnology}
		I\lowercase{ndore}, S\lowercase{imrol}, I\lowercase{ndore}, I\lowercase{ndia}.
		}

		\thanks{${}^{*}$Corresponding author.\\E-mail address: ashokiitb09@gmail.com (Ashok Kumar Pathak), 
			arshad.iitk@gmail.com (Mohd. Arshad).}

	\begin{abstract}
			{
		A new extended Farlie-Gumbel-Morgenstern copula recently studied by Ebaid et al. [Comm. Statist. Theory Methods, (2020)]  is reviewed. The reported admissible range for the copula parameter $a$ is incorrect and some typos are also found in their paper. Corrections to the admissible range of the copula parameter $a$ and typos are presented here.}
	\end{abstract}
	
	\maketitle
\section{Introduction} 
\noindent Due to simple analytical form, the family of Farlie-Gumbel-Morgenstern (FGM) copulas have been widely used in modelling dependence among random variables. Recently, a new symmetric generalization of the FGM copula has been proposed by Ebaid et al. (2020) of the form
\begin{equation}\label{CP1}
C(u,v)=uv[1+a(1-u)(1-v)(1-bu)(1-bv)],\;0\leq b\leq 2.
\end{equation}
Ebaid et al. (2020) claimed that the copula in (\ref{CP1})  achieves a higher dependence as Spearman's rho ($\rho$) and Kendall's tau ($\tau$) takes values in $[-0.33, 0.375]$ and $[-0.22, 0.25]$, respectively. However, the proposed range of the copula parameter $a$, $-1\leq a\leq \min\displaystyle \frac{1}{(1-b)},2$ has typographical error and mathematically may be express as
\begin{equation}\label{CP2}
-1\leq a\leq \min\left\{\frac{1}{1-b},2\right\}.
\end{equation} 
Unfortunately, the range (\ref{CP2}) is not correct. 
For example, if we take $b=1.1$, then using (\ref{CP2}) we get $-1\leq a\leq -10$, which does not make sense and may also give misleading values of $\rho$ and $\tau$. Here, we present a correct admissible range for the copula parameter $a$. 
\section{Correction}
\noindent It may be notice that, $C$ in (\ref{CP1}) has cubic section in both $u$ and $v$ and may be represented in the form
\begin{equation}\label{CP3}
C(u,v)=uv+a\Phi(u)\Phi(v),
\end{equation}
with $\Phi(u)=u(1-u)(1-bu)$ such that $\Phi(0)=\Phi(1)=0$  (see Rodr\'{i}guez-Lallena and \'{U}beda-Flores (2004)). The range of the copula $C$ in (\ref{CP3}) is given by 
\begin{equation*}
-{1}/{\max\{\alpha^2,\beta^2\}}\leq a\leq -{1}/{(\alpha\beta)},
\end{equation*}
where $\alpha=\inf_{u}\Phi'(u)<0$, $\beta=\sup_{u}\Phi'(u)>0$, and $\Phi'(u)$ is the first order derivative of $\Phi(u)$ with respect to $u$.\\
Let $b\in [0,2]$. After doing some calculation, we get 
\begin{equation*}
\alpha=\begin{cases}
b-1, \;\;\;\;\;\;\;\;\hfill{} \text{if}\; 0\leq b<1/2\\
\displaystyle\frac{b-b^2-1}{3b}, \; \text{if}\;   1/2\leq b\leq 2,
\end{cases}
\text{and}~~ 
\beta=1.
\end{equation*}
%
Therefore, the admissible range for the parameter $a$ for which the function $C$ in (\ref{CP1}) is a valid copula is given by
\begin{equation*}
-1\leq a \leq \begin{cases}
\displaystyle\frac{1}{(1-b)},\; \;\;\;\;\;\;\;\hfill{}\text{if}\; 0\leq b< 1/2,\\
	\displaystyle\frac{3b}{(b^2-b+1)}, \; \text{if}\; 1/2\leq b\leq 2.
\end{cases}
\end{equation*}
{\bf Remark:}
In another online version of the Ebaid et al. (2020) paper, the copula $C$ is discussed for $b\in[0,1)$. With the help of counter example, Barakat et al. (2021) showed that $C$ in (\ref{CP1}) is not a valid copula and Spearman's rho may takes values greater than unity, for the proposed admissible range $-1\leq a\leq {1}/{(1-b)}$, for $0\leq b<1$ in Ebaid et al. (2020). Unfortunately, Barakat et al. (2021) also could not identify the correct range for the parameter $a$. Let $C_{0}(u,v)=uv[1+a(1-u)(1-v)],\;-1\leq a\leq 1$ be the family of FGM copula. Using FGM copula, Barakat et al. (2021) constructed a bivariate distribution $C^{*}(u,v)=uv[1+b(1-u)][1+b(1-v)][1+a(1-u)(1-bu)(1-v)(1-bv)]$ having two different margins $G_1(u)=u[1+b(1-u)]$ and $G_2(v)=v[1+b(1-v)]$, $0\leq b<1$. They have compared copula $C$ with $C^{*}$ and  claimed that  $C(u,v)\leq C^{*}(u,v)$ for all values $0\leq u,v \leq 1$, and $0\leq b<1$. However, such type of comparison of a copula with a distribution function does not make sense. Barakat et al. (2021) also claimed that the upper limit of the maximal positive correlation $\rho$ for the copula in (\ref{CP1}) is less than 0.333. It may be noticed that, for $b=1$, (\ref{CP1})  reduces to 
$C(u,v)=uv[1+a(1-u)^2(1-v)^2]$, with maximal positive correlation $\rho=0.375$ (see Bairamov and   Bayramoglu (2013)). This  disproves Barakat et al. (2021) claim.



\end{document}